\documentclass[preprint,nofootinbib]{revtex4}%
\usepackage{amssymb}
\usepackage{amsmath}
\usepackage{amsfonts}
\usepackage{graphicx}%
\begin{document}
\begin{titlepage}
\begin{center}
\textbf{\large High-energy zero-norm states and symmetries of string
theory}
\vskip .5in
Chuan-Tsung Chan $^1$, Pei-Ming Ho $^2$, Jen-Chi Lee $^3$, \\
Shunsuke Teraguchi $^4$, Yi Yang $^3$
\vskip 15pt
{\small $^1$ Physics Division, National Center for Theoretical Sciences,
Hsinchu, Taiwan, R.O.C.} \\
{\small $^2$ Department of Physics, National Taiwan University, Taipei,
Taiwan, R.O.C.} \\
{\small $^3$ Department of Electrophysics, National Chiao-Tung University,
Hsinchu, Taiwan, R.O.C.} \\
{\small $^4$ Physics Division, National Center for Theoretical Sciences,
Taipei, Taiwan, R.O.C.}
\vskip .2in
{\small
\sffamily{ctchan@phys.cts.nthu.edu.tw} \\
\sffamily{pmho@ntu.edu.tw} \\
\sffamily{jcclee@cc.nctu.edu.tw} \\
\sffamily{teraguch@phys.ntu.edu.tw} \\
\sffamily{yiyang@mail.nctu.edu.tw}
} \\
\date{\today}
\vspace{60pt}
\textbf{abstract} \\
\end{center}
High-energy limit of zero-norm states (HZNS) in the old covariant first quantized (OCFQ)
spectrum of the 26D open bosonic string, together with the assumption
of a smooth behavior of string theory in this limit, are used to derive infinitely
many linear relations among the leading high-energy, fixed-angle behavior of
four-point functions of different string states.
As a result, ratios among all high-energy scattering amplitudes of four
\textit{arbitrary} string states can be calculated algebraically and the
leading order amplitudes can be expressed in terms of that of four tachyons
as conjectured by Gross in 1988.
A dual calculation can also be performed and equivalent results are
obtained by taking the high-energy limit of Virasoro constraints.
Finally, we compute all high-energy scattering amplitudes
of three tachyons and one massive state at the leading order
by saddle-point approximation to verify our results.
\end{titlepage}

One of the fundamental issues in string theory is its spacetime symmetry. In
local quantum field theories, a symmetry principle is usually postulated,
which can be used to fix the interactions in the theory. In string theory, on
the contrary, it is the interaction, prescribed by the very tight quantum
consistency conditions due to the extendedness of strings, which determines
the symmetry. For example, the massless gauge symmetries of 10D Heterotic
string \cite{Heter} were discovered to be $SO(32)$ or $E_{8}^{2}$ by the
string one-loop consistency or modular invariance of the theory. Symmetries
with parameters containing both Yang-Mills and Einstein indices were found
explicitly at massive levels \cite{MassHeter}. Being a consistent quantum
theory with no free parameter and an infinite number of states, it is
conceivable that there exists a huge symmetry group, which is responsible for
the ultraviolet finiteness of string theory.

Historically, a key progress to understand the symmetry of string theory is to
study the high-energy, fixed-angle behavior of string scattering amplitudes
\cite{GM, Gross, GrossManes}. This is strongly motivated by the success on the
study of the high-energy behavior of a local quantum field theory, for
example, the renormalization group and the discovery of asymptotic freedom in
QCD \cite{RG}. Moreover, the spontaneously broken symmetries are often hidden
at low energy, but become evident at high energies. Other approaches related
to this development include the studies of the high-energy, fixed momentum
transfer regime \cite{Regge}, vertex operator algebra for compactified
spacetime or on a lattice \cite{MooreWest}, the Hagedorn transition at high
temperature \cite{Hagedorn} and by taking the tensionless limit of the
worldsheet theory \cite{WS}. Despite all these efforts, a concrete picture of
the underlying symmetry of strings has not emerged.

Recently, an algebraic approach \cite{ChanLee1,ChanLee2} was developed to
derive linear relations among correlation functions for the first few massive
levels of the open bosonic string in the high-energy limit, which are
presumably a manifestation of the hidden symmetry. An important ingredient of
this approach is the zero-norm states in the old covariant first quantization
(OCFQ). In fact, it can be shown that \cite{KaoLeeCLYang} off-shell gauge
transformations of Witten string field theory, after imposing the no-ghost
condition, are identical to the on-shell stringy gauge symmetries generated by
two types of zero-norm states in the generalized massive $\sigma$-model
approach of string theory \cite{Lee}. The corresponding on-shell Ward
identities were also constructed in \cite{JCLee}. The connection between
zero-norm states and the space-time symmetry was further elucidated in the
case of 2D string where the discrete zero-norm states were shown
\cite{ChungLee} to generate a $\omega_{\infty}$ symmetry algebra
\cite{WinfinityRing}. Furthermore, these discrete zero-norm states approach to
the discrete Polyakov positive-norm states in the high-energy limit
\cite{CHLTY}.

At first sight one may raise the objection that, while the zero-norm states
correspond to gauge transformations, they can not relate physically
inequivalent states. The trick is to deform zero-norm states to some positive
norm states, which will nevertheless be called high-energy zero-norm states
(HZNS). It is the decoupling of HZNS in the high-energy limit that will allow
us to derive nontrivial relations among inequivalent amplitudes. It was
emphasized in \cite{CHL,CHLTY} that the decoupling of HZNS in the high-energy
limit is a non-trivial assumption. Those who are interested in a more detailed
account of the assumption please see \cite{CHLTY} and \cite{paperB}. Roughly
speaking, this assumption is about the regularity of the high-energy limit of
string theory.

In this letter, we will generalize the calculations for the first few mass
levels \cite{ChanLee1,ChanLee2,CHL} to arbitrary mass levels for the open
bosonic string. Infinite linear relations among high-energy scattering
amplitudes of different string states at fixed but arbitrary mass levels will
be obtained. These linear relations are then used to determine uniquely the
proportional constants among high-energy scattering amplitudes of different
string states to the leading order. Based on the high-energy scattering
amplitudes for certain representative states obtained in \cite{ChanLee1,CHL},
one can then derive the general formula of high-energy scattering amplitude
for four arbitrary string states, and express them in terms of that of four
tachyons. It was first conjectured by Gross in 1988 \cite{Gross} that
scattering amplitudes of arbitrary states are linearly related to each other
in the high-energy limit. These linear relations hint at the existence of an
infinite spacetime string symmetry hidden at low energy \cite{Gross, LEO} but
gets restored at high energies.

Let us now explain our approach in details. In the OCFQ spectrum of open
bosonic string theory, the solutions of physical state conditions include
positive-norm propagating states and two types of zero-norm states. The latter
are (we use the notation in \cite{GSW})%
\begin{align}
\text{Type I}:L_{-1}\left\vert x\right\rangle ,  &  \text{ where }%
L_{1}\left\vert x\right\rangle =L_{2}\left\vert x\right\rangle =0,\text{
}L_{0}\left\vert x\right\rangle =0;\label{1}\\
\text{Type II}:(L_{-2}+\frac{3}{2}L_{-1}^{2})\left\vert \widetilde
{x}\right\rangle ,  &  \text{ where }L_{1}\left\vert \widetilde{x}%
\right\rangle =L_{2}\left\vert \widetilde{x}\right\rangle =0,\text{ }%
(L_{0}+1)\left\vert \widetilde{x}\right\rangle =0. \label{2}%
\end{align}
While type I states have zero-norm for any space-time dimensions, type II
states have zero-norm \emph{only} for $D=26$. We take the second massive level
$M^{2}=4$ as an example to illustrate our approach. The first step is to list
the stringy Ward identities for four-point functions derived from the
decoupling of all zero-norm states at this mass level \cite{JCLee}%
\begin{align}
k_{\mu}\theta_{\nu\lambda}\mathcal{T}^{(\mu\nu\lambda)}+2\theta_{\mu\nu
}\mathcal{T}^{(\mu\nu)}  &  =0,\label{3}\\
(\frac{5}{2}k_{\mu}k_{\nu}\theta_{\lambda}^{\prime}+\eta_{\mu\nu}%
\theta_{\lambda}^{\prime})\mathcal{T}^{(\mu\nu\lambda)}+9k_{\mu}\theta_{\nu
}^{\prime}\mathcal{T}^{(\mu\nu)}+6\theta_{\mu}^{\prime}\mathcal{T}^{\mu}  &
=0,\label{4}\\
(\frac{1}{2}k_{\mu}k_{\nu}\theta_{\lambda}+2\eta_{\mu\nu}\theta_{\lambda
})\mathcal{T}^{(\mu\nu\lambda)}+9k_{\mu}\theta_{\nu}\mathcal{T} ^{[\mu\nu
]}-6\theta_{\mu}\mathcal{T}^{\mu}  &  =0,\label{5}\\
(\frac{17}{4}k_{\mu}k_{\nu}k_{\lambda}+\frac{9}{2}\eta_{\mu\nu}k_{\lambda
})\mathcal{T}^{(\mu\nu\lambda)}+(21k_{\mu}k_{\nu}+9\eta_{\mu\nu}%
)\mathcal{T}^{(\mu\nu)}+25k_{\mu}\mathcal{T}^{\mu}  &  =0, \label{6}%
\end{align}
where $\theta_{\mu\nu}$ is an arbitrary symmetric, transverse and traceless
polarization tensor, and $\theta_{\lambda}^{\prime}$ and $\theta_{\lambda}$
are arbitrary transverse vectors. $\mathcal{T}^{\prime}s$ in Eqs.(\ref{3}%
)-(\ref{6}) are four particle scattering amplitudes with, say, the second
vertex $V_{2}(k_{2})$ constructed from zero-norm states at the mass level
$M^{2}$ $=4$ and $k_{\mu}\equiv k_{2\mu}$. For example, Eq.(\ref{3}) states
the decoupling of the zero-norm state corresponding to the vertex operator
$V_{2}(k)=(k_{\mu}\theta_{\nu\lambda}\partial X^{\mu}\partial X^{\nu}\partial
X^{\lambda}+2\theta_{\mu\nu}\partial^{2}X^{\mu}\partial X^{\nu})e^{ik\cdot X}%
$, and%
\begin{equation}
\mathcal{T}^{(\mu\nu\lambda)}=\langle V_{1}\left(  \partial X^{\mu}\partial
X^{\nu}\partial X^{\lambda}e^{ik\cdot X}\right)  V_{3}V_{4}\rangle
,\qquad\mathcal{T}^{(\mu\nu)}=\langle V_{1}\left(  \partial^{2}X^{(\mu
}\partial X^{\nu)}e^{ik\cdot X}\right)  V_{3}V_{4}\rangle.
\end{equation}
The rest of the vertices $V_{1},$ $V_{3}$ and $V_{4}$ in $\mathcal{T}
^{\prime}s$ can be arbitrary string states and their tensor indices are
omitted. Note that Eqs. (\ref{3})-(\ref{6}) are valid to all loops and at all
energies $E$. We use labels 1 and 2 for incoming particles and 3 and 4 for
outgoing particles. In the center of mass frame, the scattering angle
$\phi_{c.m.}$ is defined to be the angle between $\overrightarrow{k}_{1}$ and
$\overrightarrow{k}_{3}$.

To enumerate all possible polarizations $\theta_{\mu\nu}, \theta_{\mu},
\theta^{\prime}_{\mu}$ for these equations, we define a basis of polarization
vectors for the second vertex. We have $e^{P}=\frac{1}{M_{2}}(E_{2}%
,\mathrm{k}_{2},0,\cdots)=\frac{k_{2}}{M_{2}}$ as the momentum polarization,
$e^{L}=\frac{1}{M_{2}}(\mathrm{k}_{2},E_{2},0,\cdots)$ the longitudinal
polarization and $e^{T} = (0,0,1,\cdots)$ the transverse polarization. Note
that $e^{P}$ approaches to $e^{L}$ in the high-energy limit, and the
scattering plane is defined by the spatial components of $e^{L}$ and $e^{T}$.
Polarizations perpendicular to the scattering plane are ignored because they
are kinematically suppressed for four-point scatterings in the high-energy limit.

The next step is to consider all possible polarizations for these equations
and to replace $e^{P}$ by $e^{L}$ in the high-energy limit. To the leading
order, Eqs.(\ref{3}) -(\ref{6}) imply \cite{ChanLee1,ChanLee2}%
\begin{align}
\mathcal{T}_{LLT}+\mathcal{T}_{(LT)}  &  \simeq0,\label{7}\\
10\mathcal{T}_{LLT}+\mathcal{T}_{TTT}+18\mathcal{T}_{(LT)}  &  \simeq
0,\label{8}\\
\mathcal{T}_{LLT}+\mathcal{T}_{TTT}+9\mathcal{T}_{[LT]}  &  \simeq0, \label{9}%
\end{align}
where the subscripts denote the polarizations. These linear equations can be
easily solved,%
\begin{equation}
\mathcal{T}_{TTT}:\mathcal{T}_{LLT}:\mathcal{T}_{(LT)}:\mathcal{T}%
_{[LT]}\simeq8:1:-1:-1. \label{10}%
\end{equation}
After some simple power counting of the energy order for all amplitudes
\cite{ChanLee1}, one concludes that all other amplitudes with any $V_{2}$ at
the same mass level (but with $V_{1,3,4}$ fixed) are subleading, as compared
with the amplitudes already appearing in the linear relation above. That is,
the decoupling of HZNS gives complete information about all amplitudes at the
leading order in the high-energy limit.

This linear relation among scattering amplitudes agrees with direct
computation. For example, consider the four-point functions with one tensor
$V_{2}$ and three tachyons $V_{1,3,4}$. The $s-t$ channel contributions to the
scattering amplitudes in Eq.(\ref{10}) are given by \cite{ChanLee1,ChanLee2}
$\mathcal{T}_{TTT}\simeq-8E^{9}\sin^{3}\phi_{c.m.}\mathcal{T}(3)\simeq
8\mathcal{T}_{LLT}\simeq-8\mathcal{T}_{(LT)}\simeq-8\mathcal{T}_{[LT]}$, where
$\mathcal{T}(3)$ is given by%
\begin{align}
\mathcal{T}(n)  &  =\sqrt{\pi}(-1)^{n-1}2^{-n}E^{-1-2n}(\sin\frac{\phi_{c.m.}%
}{2})^{-3}(\cos\frac{\phi_{c.m.}}{2})^{5-2n}\nonumber\\
&  \times\exp(-\frac{s\ln s+t\ln t-(s+t)\ln(s+t)}{2}), \label{11}%
\end{align}
which is the high-energy limit of $\frac{\Gamma(-\frac{s}{2}-1)\Gamma
(-\frac{t}{2}-1)}{\Gamma(\frac{u}{2}+2)}$ with $s+t+u=2n-8$. Here
$s=-(k_{1}+k_{2})^{2}$, $t=-(k_{2}+k_{3})^{2}$ and $u=-(k_{1}+k_{3})^{2}$ are
the Mandelstam variables.

%In deriving Eqs.(\ref{7}) - (\ref{9}), we replaced the polarization $e^P$
%by $e^L$ in the correlation functions
%and assumed that this is a good approximation in the high-energy limit.
%We will make this assumption throughout the paper as there is no
%violation of it in all cases that have been considered.
%A similar calculation for the mass level $M^{2} = 6$ was carried out in \cite{ChanLee2}.

The aim of this letter is to show that there is indeed only one independent
component of high-energy scattering amplitude at the leading order for general
mass levels, and to calculate the proportional constants generalizing
Eq.(\ref{10}). One first notes that the decoupling of HZNS indicates that the
only states that will survive the high-energy limit at level $M^{2}=2(n-1)$
are of the form%
\begin{equation}
\left\vert n,2m,q\right\rangle \equiv(\alpha_{-1}^{T})^{n-2m-2q}(\alpha
_{-1}^{L})^{2m}(\alpha_{-2}^{L})^{q}\left\vert 0,k\right\rangle .\label{13}%
\end{equation}
It is algebraically proved that an amplitude becomes subleading if we replace
a state of this form by another state not of this form at the same mass level
in \cite{CHLTY,paperB}. One can also verify this fact by explicitly
calculating an amplitude using saddle point method \cite{CHLTY,CHL}. The
reason is as follows \ First of all, polarizations in $P$ can always be gauged
away such that the polarizations of positive-norm physical states are given in
terms of $L$ and $T$ only. Secondly, each factor of $\alpha_{-1}^{T}$
corresponds to a factor of $\partial X^{T}$ in the vertex. In the high-energy
limit, its leading order contribution comes from its contraction with the
exponent of $e^{ik\cdot X}$ in another vertex. Hence each factor of
$\alpha_{-1}^{T}$ contributes a factor of $k\propto E$ in the high-energy
limit. The operator $(\alpha_{-1}^{L})^{2m}$, on the other hand, contributes
energy order $E^{2m}$. Naively, each factor of $\alpha_{-1}^{L}$ contributes a
factor of $e^{L}\cdot k\propto E^{2}$ due to the contraction of $\partial
X^{L}$ with the exponent of $e^{ik\cdot X}$ in another vertex. However, in the
saddle-point computation, the leading order amplitude will drop $m$ times from
$E^{4m}$ to $E^{2m}$, since the leading contribution associated with
$(\alpha_{-1}^{L})^{2m}$ comes from the higher order ($\frac{1}{\alpha
^{\prime}}$) expansion, see Eq.(22). Similarly, the operator $(\alpha_{-1}%
^{L})^{2m-1}$ is of energy order $E^{2m-2}$, and can be neglected at
high-energies when compared with $(\alpha_{-1}^{T})^{2m-1}$. (Since the factor
$(\alpha_{-1}^{L})^{2m-1}$ in a state can be replaced by $(\alpha_{-1}%
^{T})^{2m-1}$ without changing the mass level, the former is subleading.) This
is why we only keep even powers of $\alpha_{-1}^{L}$ in Eq.(\ref{13}).
Finally, the operator $\alpha_{-2}^{T}$ is of energy order $E$ but carries
mass unit two, so it is subleading when compared with $(\alpha_{-1}^{T})^{2}$.
Similarly, since the energy order of the operator $\alpha_{-m}^{\mu}$ is no
greater than $E^{2}$, those operators with $m\geq3$ are subleading and can be
neglected. A more rigorous proof of these arguments can be found in
\cite{CHLTY,paperB}.

The next step is to use the decoupling of two types of HZNS%
\begin{align}
L_{-1}\left\vert n-1,2m-1,q\right\rangle  &  \simeq M\left\vert
n,2m,q\right\rangle +(2m-1)\left\vert n,2m-2,q+1\right\rangle ,\label{14}\\
L_{-2}\left\vert n-2,0,q\right\rangle  &  \simeq\frac{1}{2}\left\vert
n,0,q\right\rangle +M\left\vert n,0,q+1\right\rangle , \label{15}%
\end{align}
to deduce the ratios of all amplitudes at the leading order. The final result
is \cite{CHLTY}%
\begin{equation}
\mathcal{T}^{(n,2m,q)}=\left(  -\frac{1}{M}\right)  ^{2m+q}\left(  \frac{1}%
{2}\right)  ^{m+q}(2m-1)!!\mathcal{T}^{(n,0,0)}. \label{16}%
\end{equation}
Eq.(\ref{16}) correctly reproduces the proportional constants for mass level
$M^{2}=4,6$ after the Young tableaux decomposition \cite{ChanLee1,ChanLee2}.

Eq.(\ref{16}) also allows us to obtain the general formula for four particle
scattering amplitudes at the tree level in the high-energy limit%
\begin{equation}
\langle V_{1}V_{2}V_{3}V_{4}\rangle={\textstyle\prod_{i=1}^{4}}\left(
-\frac{1}{M_{i}}\right)  ^{2m_{i}+q_{i}}\left(  \frac{1}{2}\right)
^{m_{i}+q_{i}}(2m_{i}-1)!!\mathcal{T}_{n_{1}n_{2}n_{3}n_{4}}^{T^{1}\cdot\cdot
T^{2}\cdot\cdot T^{3}\cdot\cdot T^{4}\cdot\cdot}, \label{17}%
\end{equation}
where $\mathcal{T}_{n_{1}n_{2}n_{3}n_{4}}^{T^{1}\cdot\cdot T^{2}\cdot\cdot
T^{3}\cdot\cdot T^{4}\cdot\cdot}$ is the high-energy scattering amplitude for
$V_{i}=(e^{T^{i}}\cdot\partial X)^{n_{i}}e^{ik_{i}X_{i}}$ $(i=1,\cdot
\cdot\cdot,4)$. ($T^{i}$ is the transverse polarization for the $i$-th
particle. $e^{T}$ defined earlier is denoted $e^{T^{2}}$ here.) It only
depends on the sum $n=\sum_{i=1}^{4}n_{i}$ and is given by \cite{ChanLee1,CHL}%
\begin{equation}
\mathcal{T}_{n_{1}n_{2}n_{3}n_{4}}^{T^{1}\cdot\cdot T^{2}\cdot\cdot T^{3}%
\cdot\cdot T^{4}\cdot\cdot}=[-2E^{3}\sin\phi_{c.m.}]^{\Sigma n_{i}}%
\mathcal{T}(\Sigma n_{i})=2\sqrt{\pi}e^{n-4}(stu)^{\frac{n-3}{2}}e^{-\frac
{1}{2}(s\ln s+t\ln t+u\ln u)}. \label{12}%
\end{equation}

It is interesting to see that the contribution of the second term $\frac{3}{2}
L_{-1}^{2}\left\vert \widetilde{x}\right\rangle $ of type II zero-norm states
in Eq.(\ref{2}) to the stringy Ward identities is negligible in the
high-energy limit. This hints at a \textquotedblleft dual\textquotedblright%
\ calculation, the Virasoro constraints, to derive Eq.(\ref{16}) \cite{CHLTY}.
It is convenient to use Young tableaux to handle the complicated symmetric
structures of tensor fields with higher spins. The most general state in the
mass level $M^{2}=2\left(  n-1\right)  $ can be written as%
\begin{equation}
\left\vert n\right\rangle =\left\{  \sum_{m_{j}}\overset{k}{\underset
{j=1}{\otimes}}\frac{1}{j^{m_{j}}m_{j}!}%
\begin{tabular}
[c]{|c|c|c|}\hline
$\mu_{1}^{j}$ & $\cdots$ & $\mu_{m_{j}}^{j}$\\\hline
\end{tabular}
\ \ \alpha_{-j}^{\mu_{1}^{j}}\cdots\alpha_{-j}^{\mu_{m_{j}}^{j}}\right\}
\left\vert 0,k\right\rangle , \label{general state}%
\end{equation}
where $m_{j}$ is the number of the operator $\alpha_{-j}^{\mu}$ and the
summation runs over all possible combinations of $m_{j}$ with $\sum_{j=1}%
^{k}jm_{j}=n$. Since the upper indices $\left\{  \mu_{1}^{j}\cdots\mu_{m_{j}%
}^{j}\right\}  $ for $\alpha_{-j}^{\mu_{1}^{j}}\cdots\alpha_{-j}^{\mu_{m_{j}%
}^{j}}$ are symmetric, we can use the \emph{1-row} Young tableaux to stand for
the coefficients in (\ref{general state}). The direct product $\otimes$ acts
on the Young tableaux in the standard way. Next, we will apply the Virasoro
constraints to the state (\ref{general state}). The only Virasoro constraints
which need to be considered are $L_{1}\left\vert n\right\rangle =L_{2}%
\left\vert n\right\rangle =0$. They give, in the high-energy limit
\cite{CHLTY},%
%TCIMACRO{\TeXButton{%
%\begin{subequations}%
%}{\begin{subequations}}}%
%BeginExpansion
\begin{subequations}%
%EndExpansion%
\begin{align}
\underset{n-2q-2m}{\underbrace{%
\begin{tabular}
[c]{|l|lll|l|}\hline
$T$ &  & $\cdots$ &  & $T$\\\hline
\end{tabular}
\ \ }}\underset{2m}{\underbrace{%
\begin{tabular}
[c]{|l|l|l|}\hline
$L$ & $\cdots$ & $L$\\\hline
\end{tabular}
\ \ }}\otimes\underset{q}{\underbrace{%
\begin{tabular}
[c]{|l|l|l|}\hline
$L$ & $\cdots$ & $L$\\\hline
\end{tabular}
\ \ }}  &  =-\frac{2m-1}{M}\underset{n-2q-2m}{\underbrace{%
\begin{tabular}
[c]{|l|lll|l|}\hline
$T$ &  & $\cdots$ &  & $T$\\\hline
\end{tabular}
\ \ }}\underset{2m-2}{\underbrace{%
\begin{tabular}
[c]{|l|l|l|}\hline
$L$ & $\cdots$ & $L$\\\hline
\end{tabular}
\ \ }}\otimes\underset{q+1}{\underbrace{%
\begin{tabular}
[c]{|l|l|l|}\hline
$L$ & $\cdots$ & $L$\\\hline
\end{tabular}
\ \ }},\label{L1-}\\
\underset{n-2q-2-2m}{\underbrace{%
\begin{tabular}
[c]{|l|lll|l|}\hline
$T$ &  & $\cdots$ &  & $T$\\\hline
\end{tabular}
\ \ }}\underset{2m}{\underbrace{%
\begin{tabular}
[c]{|l|l|l|}\hline
$L$ & $\cdots$ & $L$\\\hline
\end{tabular}
\ \ }}\otimes\underset{q+1}{\underbrace{%
\begin{tabular}
[c]{|l|l|l|}\hline
$L$ & $\cdots$ & $L$\\\hline
\end{tabular}
\ \ }}  &  =-\frac{1}{2M}\underset{n-2q-2m}{\underbrace{%
\begin{tabular}
[c]{|l|lll|l|}\hline
$T$ &  & $\cdots$ &  & $T$\\\hline
\end{tabular}
\ \ }}\underset{2m}{\underbrace{%
\begin{tabular}
[c]{|l|l|l|}\hline
$L$ & $\cdots$ & $L$\\\hline
\end{tabular}
\ \ }}\otimes\underset{q}{\underbrace{%
\begin{tabular}
[c]{|l|l|l|}\hline
$L$ & $\cdots$ & $L$\\\hline
\end{tabular}
\ \ }}. \label{L2-}%
\end{align}%
%TCIMACRO{\TeXButton{%
%\end{subequations}%
%}{\end{subequations}} }%
%BeginExpansion
\end{subequations}
%EndExpansion
It is worth emphasizing that these equations are equivalent to Eqs.(14) and
(15) in the first approach. Combining Eqs.(\ref{L1-}) and (\ref{L2-}) gives
our main result Eq.(\ref{16}).

Finally, we compute all high-energy scattering amplitudes of three tachyons
and one massive state Eq.(\ref{13}) by saddle-point approximation \cite{CHL}
to justify our result Eq.(\ref{16}). The $s-t$ channel contribution to the
high-energy scattering amplitude at tree level is%
\begin{align}
\mathcal{T}^{(n,2m,q)}  &  \simeq\int_{0}^{1}dx\;x^{k_{1}.k_{2}}%
(1-x)^{k_{2}.k_{3}}\left[  \frac{e^{T}\cdot k_{1}}{x}-\frac{e^{T}\cdot k_{3}%
}{1-x}\right]  ^{n-2m-2q}\nonumber\\
&  \cdot\left[  \frac{e^{P}\cdot k_{1}}{x}-\frac{e^{P}\cdot k_{3}}%
{1-x}\right]  ^{2m}\left[  -\frac{e^{P}\cdot k_{1}}{x^{2}}-\frac{e^{P}\cdot
k_{3}}{(1-x)^{2}}\right]  ^{q}, \label{singletensor}%
\end{align}
where we have substituted the polarization $L$ by $P$. In order to apply the
saddle-point method, we rewrite the amplitude above in the form $\mathcal{T}%
^{(n,2m,q)}(K)=\int_{0}^{1}dx$ $u(x)e^{-Kf(x)},$ where $K\equiv-k_{1}%
.k_{2}\rightarrow2E^{2},$ $f(x)\equiv\ln x-\tau\ln(1-x)$ $,\tau\equiv
-\frac{k_{2}.k_{3}}{k_{1}.k_{2}}\rightarrow\sin^{2}\frac{\phi_{c.m.}}{2}$ and
$u(x)\equiv\left[  \frac{k_{1}.k_{2}}{M}\right]  ^{q+2m}(1-x)^{-n+2m+2q}%
(f^{\prime})^{2m}(f^{\prime\prime})^{q}(-e^{T}\cdot k_{3})^{n-2m-2q}$. The
saddle-point for the integration of moduli, $x=x_{0}=\frac{1}{1-\tau}$, is
defined by $f^{\prime}(x_{0})=0$.{}From the definition of $u(x)$, it is easy
to see that $u(x_{0})=u^{\prime}(x_{0})=....=u^{(2m-1)}(x_{0})=0,$
\cite{footnote} and one can easily compute $u^{(2m)}(x_{0})$ and evaluate the
Gaussian integral%
\begin{align}
\mathcal{T}^{(n,2m,q)}  &  =\int_{0}^{1}dx\text{ }u(x)e^{-Kf(x)}=\sqrt
{\frac{2\pi}{Kf_{0}^{\prime\prime}}}e^{-Kf_{0}}\left[  \frac{u_{0}^{(2m)}%
}{2^{k}\ m!\ (f_{0}^{\prime\prime})^{m}\ K^{m}}+O(\frac{1}{K^{m+1}})\right]
\nonumber\\
&  =\sqrt{\frac{2\pi}{Kf_{0}^{\prime\prime}}}e^{-Kf_{0}}\left[  (-1)^{n-m}%
\frac{2^{n-q-2m}(2m)!}{k!\ {M}^{q+2m}}\ \tau^{-\frac{n}{2}}(1-\tau)^{\frac
{3n}{2}}E^{n}+O(E^{n-2})\right]  . \label{leading}%
\end{align}
This result shows that the high-energy four-point amplitudes of states at
fixed mass level $n$ share the same energy and angular dependence, and one can
compute the ratios among high-energy amplitudes to correctly reproduce
Eq.(\ref{16}). More details can be found in \cite{CHLTY}.

Although our discussions focus on the 26D open bosonic string theory, the same
approach can be applied to superstrings as well \cite{susy}. The infinitely
many linear relations which uniquely determine the ratios of all high-energy
four-point functions at a fixed mass level strongly suggest the hidden
symmetry of the string theory. Even though we have not identified the symmetry
group for the string theory, this work sheds new light towards the final answer.

We thank Jiunn-Wei Chen, Tohru Eguchi, Koji Hashimoto, Hiroyuki Hata, Takeo
Inami, Hsien-Chung Kao, Yeong-Chuan Kao, Yoichi Kazama, Yutaka Matsuo and
Tamiaki Yoneya for discussions. This work is supported in part by the National
Science Council, Taiwan, R.O.C and National Center for Theoretical Sciences,
Hsinchu, Taiwan, R.O.C (grant NSC 94-2119-M-002-001).


\begin{thebibliography}{99}                                                                                               %


\bibitem {Heter}D.~J.~Gross, J.~A.~Harvey, E.~J.~Martinec and R.~Rohm,
Nucl.\ Phys.\ B \textbf{256}, 253 (1985); Nucl.\ Phys.\ B \textbf{267}, 75 (1986).

\bibitem {MassHeter}J.~C.~Lee, Phys.\ Lett.\ B \textbf{337}, 69 (1994).

\bibitem {GM}D.~J.~Gross and P.~F.~Mende, Phys.\ Lett.\ B \textbf{197}, 129
(1987); Nucl.\ Phys.\ B \textbf{303}, 407 (1988).

\bibitem {Gross}D.~J.~Gross, Phys.\ Rev.\ Lett.\ \textbf{60}, 1229 (1988);
Phil.\ Trans.\ R. Soc. Lond. A329, 401 (1989).

\bibitem {GrossManes}D.~J.~Gross and J.~L.~Manes, Nucl.\ Phys.\ B
\textbf{326}, 73 (1989). See section 6 for details.

\bibitem {RG}D.~J.~Gross and F.~Wilczek, Phys.\ Rev.\ Lett.\ \textbf{30}, 1343
(1973); Phys.\ Rev.\ D \textbf{8}, 3633 (1973); Phys.\ Rev.\ D \textbf{9}, 980
(1974). H.~D.~Politzer, Phys.\ Rev.\ Lett.\ \textbf{30}, 1346 (1973).

\bibitem {Regge}D.~Amati, M.~Ciafaloni and G.~Veneziano, Phys.\ Lett.\ B
\textbf{216}, 41 (1989).

\bibitem {MooreWest}G.~W.~Moore, [arXiv:hep-th/9305139]; G.~W.~Moore,
[arXiv:hep-th/9310026]. P.~C.~West, Mod.\ Phys.\ Lett.\ A \textbf{10}, 761
(1995). N.~Moeller and P.~West, Nucl.\ Phys.\ B \textbf{729}, 1 (2005).

\bibitem {Hagedorn}J.~J.~Atick and E.~Witten, Nucl.\ Phys.\ B \textbf{310},
291 (1988). R.~Hagedorn, Nuovo Cim.\ Suppl.\ \textbf{3}, 147 (1965).

\bibitem {WS}See, for instance, J.~Isberg, U.~Lindstrom, B.~Sundborg and
G.~Theodoridis, Nucl.\ Phys.\ B \textbf{411}, 122 (1994). B.~Sundborg,
Nucl.\ Phys.\ Proc.\ Suppl.\ \textbf{102}, 113 (2001). E.~Sezgin and
P.~Sundell,\ Nucl.\ Phys.\ B \textbf{644}, 303 (2002) [Erratum-ibid.\ B
\textbf{660}, 403 (2003)]. C.~S.~Chu, P.~M.~Ho and F.~L.~Lin, JHEP
\textbf{0209}, 003 (2002) G.~Bonelli,\ Nucl.\ Phys.\ B \textbf{669}, 159 (2003).

\bibitem {ChanLee1}C.~T.~Chan and J.~C.~Lee, Phys.\ Lett.\ B \textbf{611}, 193
(2005). J.~C.~Lee, . [arXiv:hep-th/0303012].

\bibitem {ChanLee2}C.~T.~Chan and J.~C.~Lee, Nucl.\ Phys.\ B \textbf{690}, 3 (2004).

\bibitem {KaoLeeCLYang}H.~C.~Kao and J.~C.~Lee, Phys.\ Rev.\ D \textbf{67},
086003 (2003). C.~T.~Chan, J.~C.~Lee and Y.~Yang, Phys.\ Rev.\ D \textbf{71},
086005 (2005)

\bibitem {Lee}J.~C.~Lee, Phys.\ Lett.\ B \textbf{241}, 336 (1990);
Phys.\ Rev.\ Lett.\ \textbf{64}, 1636 (1990). J.~C.~Lee and B.~Ovrut,
Nucl.\ Phys.\ B \textbf{336}, 222 (1990).

\bibitem {JCLee}J.~C.~Lee, Prog.\ Theor.\ Phys.\ \textbf{91}, 353 (1994).

\bibitem {ChungLee}T.~D.~Chung and J.~C.~Lee, Phys.\ Lett.\ B \textbf{350}, 22
(1995) ; \ Z.\ Phys.\ C \textbf{75}, 555 (1997).
J.~C.~Lee,\ Eur.\ Phys.\ J.\ C \textbf{1}, 739 (1998).

\bibitem {WinfinityRing}J.~Avan and A.~Jevicki, Phys.\ Lett.\ B \textbf{266},
35 (1991); Phys.\ Lett.\ B \textbf{272}, 17 (1991). I.~R.~Klebanov and
A.~M.~Polyakov, Mod.\ Phys.\ Lett.\ A \textbf{6}, 3273 (1991). E.~Witten,
Nucl.\ Phys.\ B \textbf{373}, 187 (1992). E.~Witten and B.~Zwiebach,
Nucl.\ Phys.\ B \textbf{377}, 55 (1992).

\bibitem {CHLTY}C.~T.~Chan, P.~M.~Ho, J.~C.~Lee, S.~Teraguchi and Y.~Yang,
Nucl.\ Phys.\ B \textbf{725}, 352 (2005).

\bibitem {CHL}C.~T.~Chan, P.~M.~Ho and J.~C.~Lee, Nucl.\ Phys.\ B
\textbf{708}, 99 (2005).

\bibitem {paperB}C.~T.~Chan, P.~M.~Ho, J.~C.~Lee, S.~Teraguchi and Y.~Yang,
\textquotedblleft Comments on the high energy limit of bosonic open string
theory,\textquotedblright\ [arXiv:hep-th/0509009].

\bibitem {LEO}J.~C.~Lee, Phys.\ Lett.\ B \textbf{326}, 79 (1994). M.~Evans and
B.~A.~Ovrut, Phys.\ Lett.\ B \textbf{231}, 80 (1989).

\bibitem {GSW}M.~B.~Green, J.~H.~Schwarz and E.~Witten, ``Superstring Theory.
Vol. 1,'' Cambridge University Press 1987.

\bibitem {footnote}Notice that we need to take the $2m$-th order saddle point
approximation in order to find the leading order coefficient of the
correlation function. This explains \cite{CHL} what is missing in the previous
calculations \cite{GM,Gross,GrossManes}.

\bibitem {susy}C.~T.~Chan, J.~C.~Lee and Y.~Yang, Nucl.\ Phys.\ B
\textbf{738}, 93 (2006).
\end{thebibliography}
\end{document}